\documentclass[ ] {aipproc}
\layoutstyle{6x9}

\begin{document}
\title{Fresnel zone plates for Achromatic Imaging Survey of X-ray sources}
\author{Sourav Palit}{address={Indian Centre for Space Physics (ICSP),
Chalantika 43, Garia Station Rd., Kolkata, 700084}}
\author{S. K. Chakrabarti}{address={S.N. Bose National Center for Basic
Sciences, JD-Block, Salt Lake, Kolkata, 700098}
,altaddress={ICSP (Principal Investigator); chakraba@bose.res.in}}
\author{D. Debnath}{address={Indian Centre for Space Physics (ICSP),
Chalantika 43, Garia Station Road, Kolkata 700084}}
\author{Vipin Yadav}{address={Indian Centre for Space Physics (ICSP),
Chalantika 43, Garia Station Road, Kolkata 700084}
,altaddress={On deputation from Indian Space Research Organization HQ,
Bangalore 560 231}}
\author{Anuj Nandi}{address={Indian Centre for Space Physics (ICSP),
Chalantika 43, Garia Station Road, Kolkata 700084}
,altaddress={On deputation from Indian Space Research Organization HQ,
Bangalore 560 231}}

\begin{abstract}

\noindent
A telescope with Fresnel Zone Plates has been contemplated to be an excellent imaging
mask in X-rays and gamma-rays for quite some time. With a proper choice
of zone plate material, spacing and an appropriate readout system it is
possible to achieve any theoretical angular resolution. We provide the
results of numerical simulations of how a large number of X-ray sources
could be imaged at a high resolution. We believe that such an imager
would be an excellent tool for a future survey mission for X-ray and
gamma-ray sources which we propose.

\end{abstract}

\maketitle
%%%%%%%%%%%%%%%%%%%%%%%%%%%%%%%%%%%%%%%%%%%%%%%%%%%%%%%%%%%%%%%%%%%%%%%
\section{Introduction}

\noindent
The usage of Fresnel Zone plate combinations have been suggested to be
an excellent way to obtain a very high resolution imaging in X-rays and
gamma-rays \cite{Mertz1961, Mertz1965} and several workers, specifically
Desai and collaborators \cite{Desai1991, Desai1998, Desai1999} have investigated 
this possibility for over a decade.

\noindent
The advantage of the zone plate telescope is that, in principle, an
arbitrarily high resolution can be achieved by it. The angular
resolution is $\delta\theta = \delta y / D$ where $\delta y$ is the
finest zone at the outer boundary and $D$ is the distance between two
plates. The plates can be made of materials which is opaque to the
observational band-width and it will remain achromatic in the entire
range of interest. For instance, a single plate of 1 mm will block about
$\sim 150 ~keV$ photons and with a pair which will cast a shadow on the
detector system one would achieve the identical angular resolution even
at $250 ~keV$ without any problem. The major problem appears to be the
pointing accuracy of a detector system. In fact, with a pair of zone
plates of $\sim 30 ~mm$ diameter with $50 \mu m$ outer zone at a
distance of $10 m$, one can have a resolution of about an arc second. This
is achievable. With the advent of nano technology, one can even
conceive of a few $nm$ of the zone width and achievements of a
mili-arcsecond of resolution, extending the possibility of directly
imaging the immediate vicinity of a super-massive black hole. By
increasing the distance between the plates (as in two satellites) even
higher resolution (e.g., micro arc seconds) could be achieved.
Fairly close binary X-ray sources (such as WW Aurigae) could also 
be resolved easily.

\noindent
While achieving achromatic images over a large energy range, the
thickness of a plate has to be large. As we mentioned, for a tungsten
plate, to have an image at $150 ~keV$, we need to have the
plate-thickness of $d = 1 ~mm$. Similarly, to achieve high resolution,
we require $\delta y$ to be low. This limits the highest possible angle
($\delta y/d$ rad for $\delta y = 50 ~\mu m$ and $ d = 0.1 ~cm$) of the
source whose photons can enter through the transparent zone. Thus, the
survey will have to be made from a narrow field collimator. In order to
achieve statistically significant photon count, survey has to be done
with a mosaic mode and not by continuously scanning the sky.

\begin{figure}[h]
\includegraphics[height=6cm]{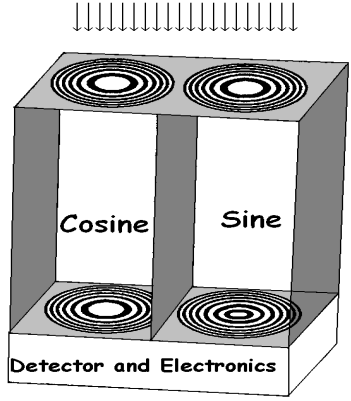}
\caption{Schematic diagram of the experimental set based on which the
simulation is made. One set of Cosine and one set of Sine combinations
are kept side by side. In the detector box, we assume a CMOS CCD detector 
of $50 ~\mu m$ pixel width.} 
\end{figure}

\noindent
In the present paper, we propose a survey mission of the sky at wide
band of X-rays energies (i.e., $\sim 1-250 ~keV$ range)
with a moderately high resolution. In Fig. 1, we show a typical 
zoneplate telescope with two pairs of zone
plates, one pair is of {\it Sine} type and the other pair is of 
{\it Cosine} type. Both are positive (i.e., the
central zones are transparent to X-rays). We present simulation results
which show distribution of photons on the detector plane. We then carry
out Inverse Fourier Transformation to retrieve the source distribution.
Cosine combination of zone plates consists of zones whose radii vary as 
$\sqrt n$ and Sine combination consists of zones whose radii vary as $\sqrt {n-1/2}$. 

%%%%%%%%%%%%%%%%%%%%%%%%%%%%%%%%%%%%%%%%%%%%%%%%%%%%%%%%%%%%%%%%%%%%%%%
\section{Simulations results}

\noindent
For the sake of concreteness, we concentrate on the simulation of cases
with only two sources and twelve sources in the field ($2^\circ \times 2^\circ$).
We assume zone plates of $34 mm$ diameter and 50 $\mu m$ as the smallest
zone thickness. The tungsten made plates are of $1 mm$ thick and they are separated by
$30 cm$. A CMOS CCD ($1024 \times 1024$ pixels with 1 square pixel = 50
$\mu m ^2$) is assumed to act as the detector. $\theta$ in degrees (measured from
the vertical axis clockwise) and $\phi$ in arc seconds (measured
from the axis of symmetry of the zone plate) define the nature of
off-axisness of the source. 

\begin{figure}
\includegraphics[height=5cm]{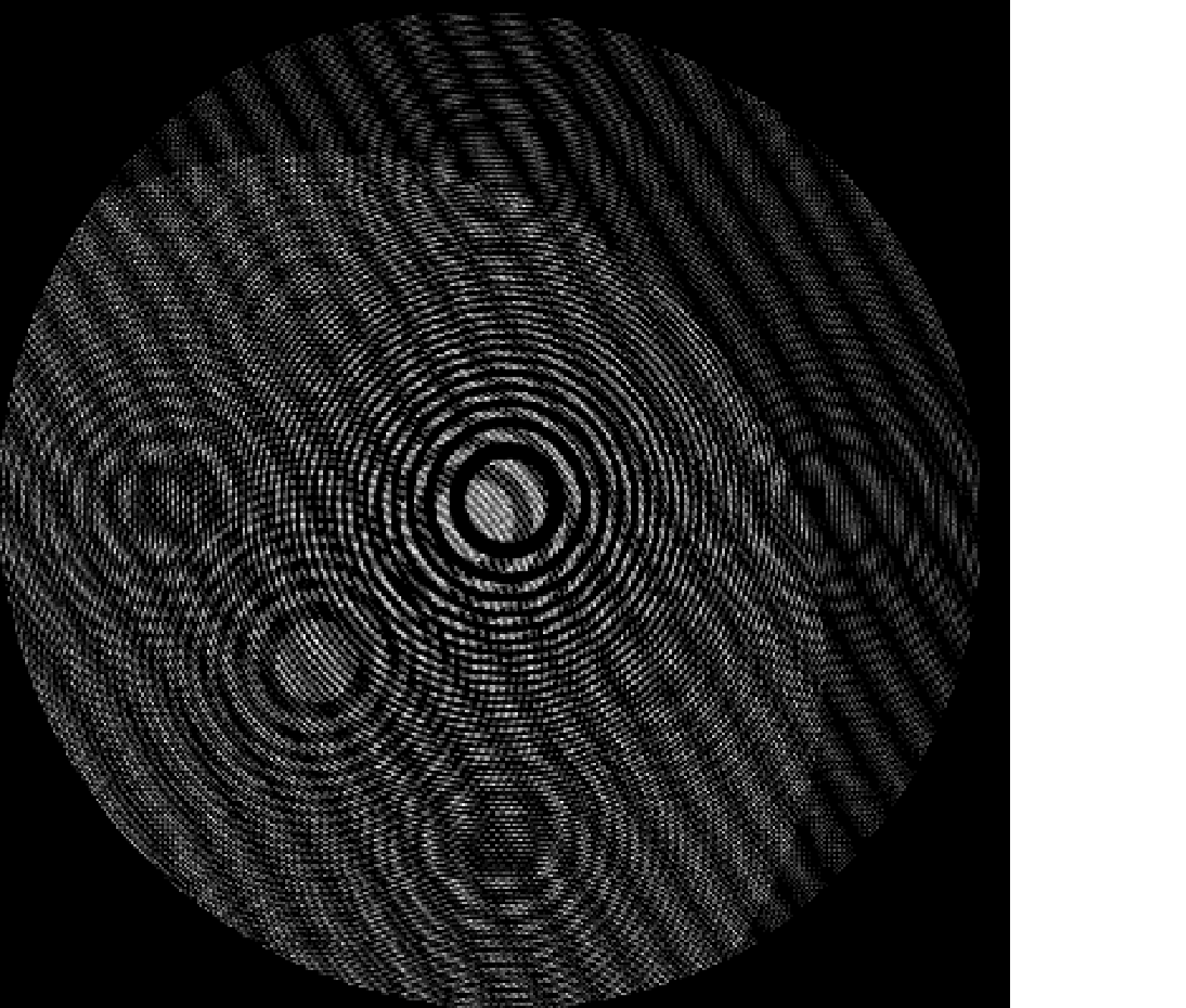}
\includegraphics[height=7cm]{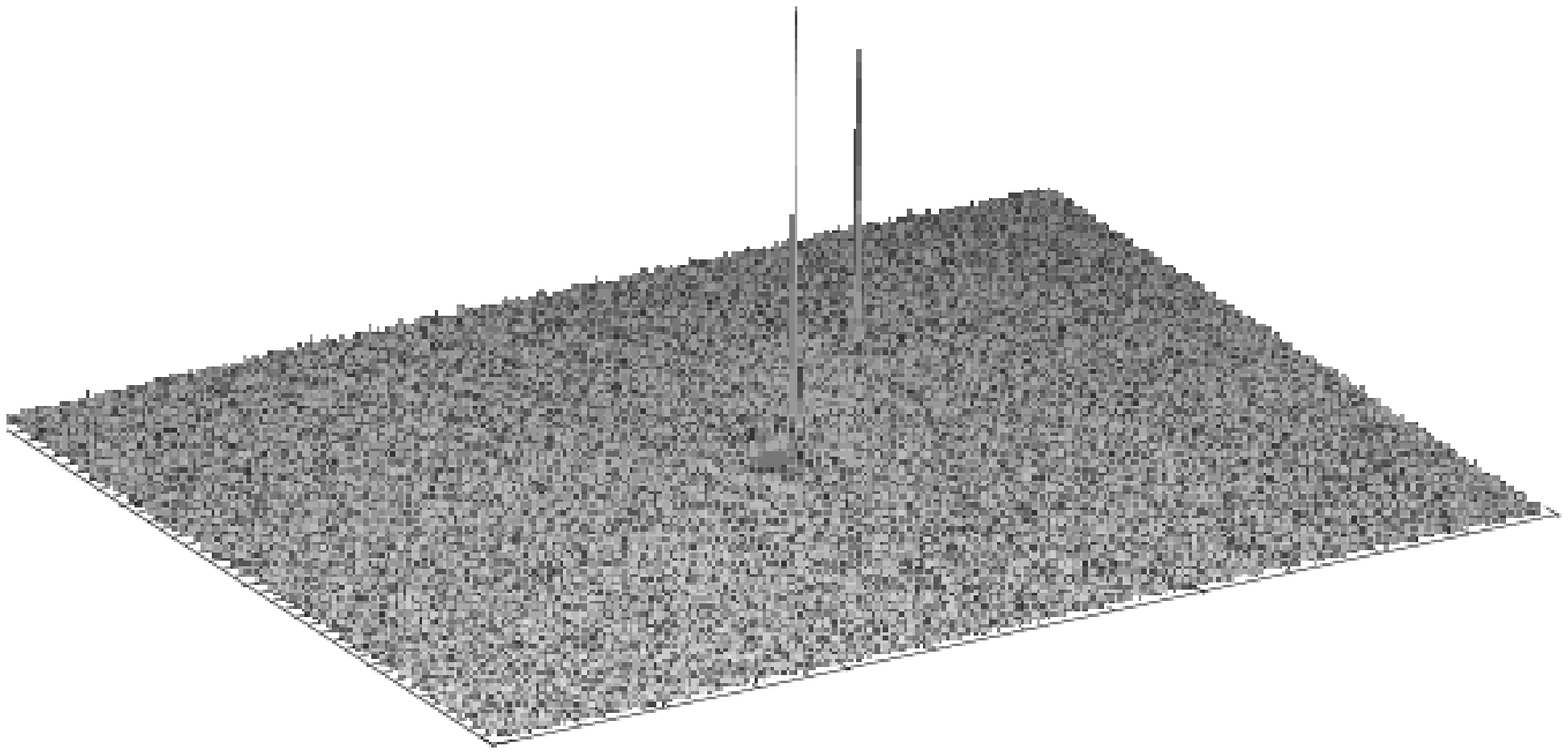}
\caption{(a) Simulated Moire Fringe patterns on the detector plane when two
sources are placed at $800$ arcsec and $5000$ arcsec off axis locations.
(b) Three dimensional reconstruction of the photon counts with the base FOV 
$2^\circ \times 2^\circ$.}
\end{figure}

\begin{figure}
\includegraphics[height=5cm]{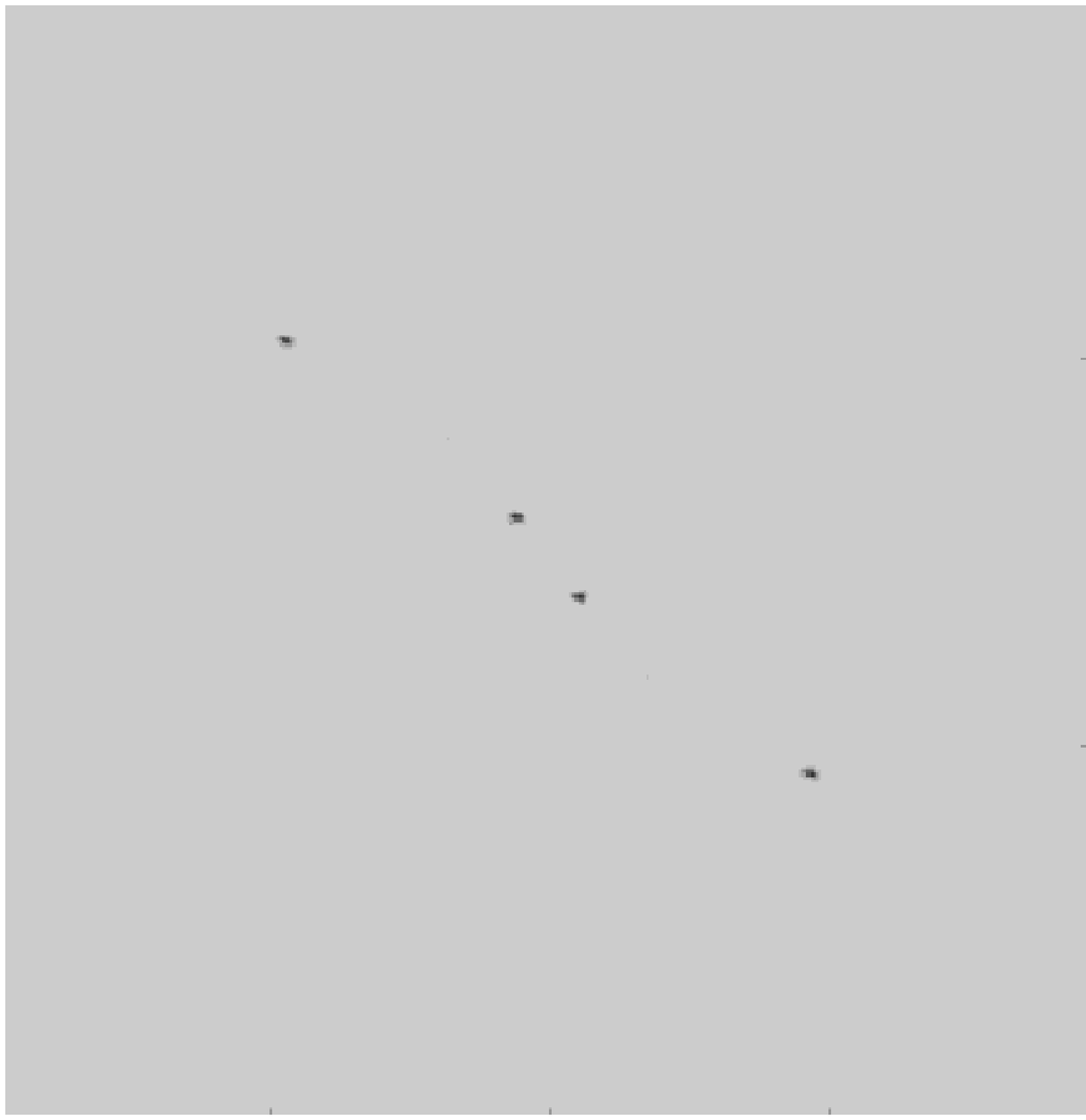}
\hspace{20mm}
\includegraphics[height=5cm]{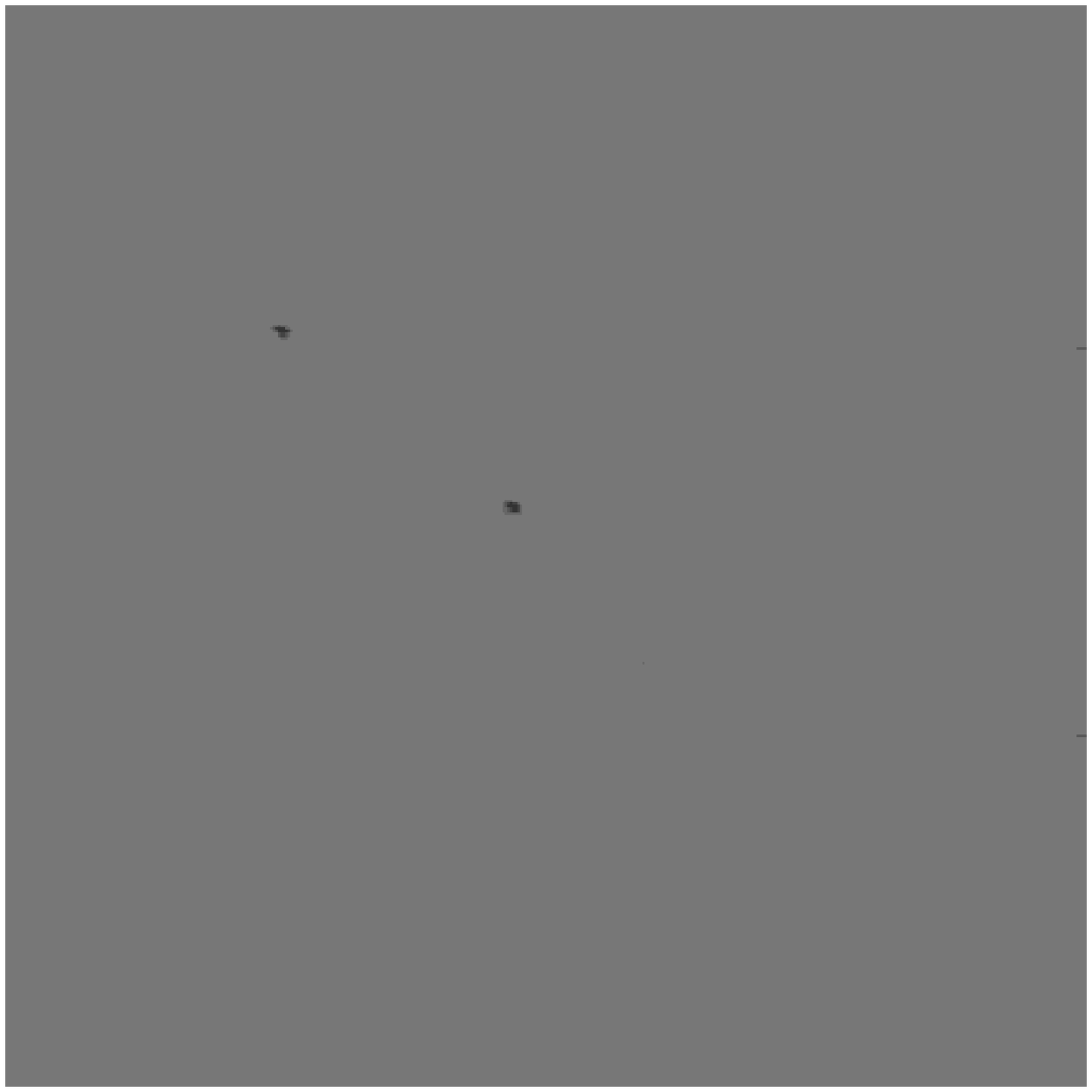}
\caption{(a) Reconstruction of the two sources from Fig. 2a using
only the Cosine transformation. Note that there are two sets of sources
symmetrically placed. (b) Same as in (a) except
using both the Cosine and Sine transformations.}
\end{figure}

\begin{figure}
\includegraphics[height=5cm]{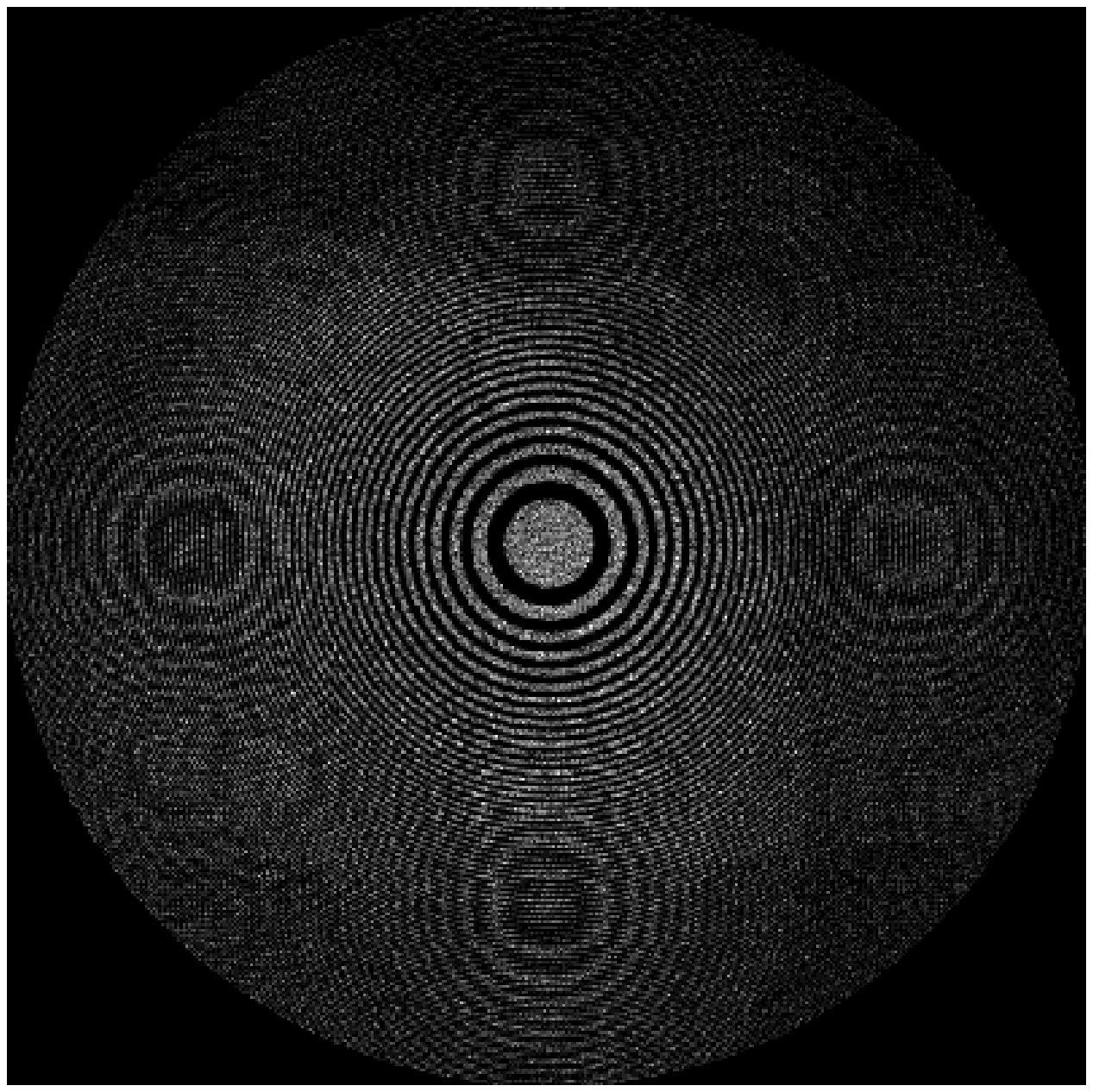}
\includegraphics[height=7.0cm]{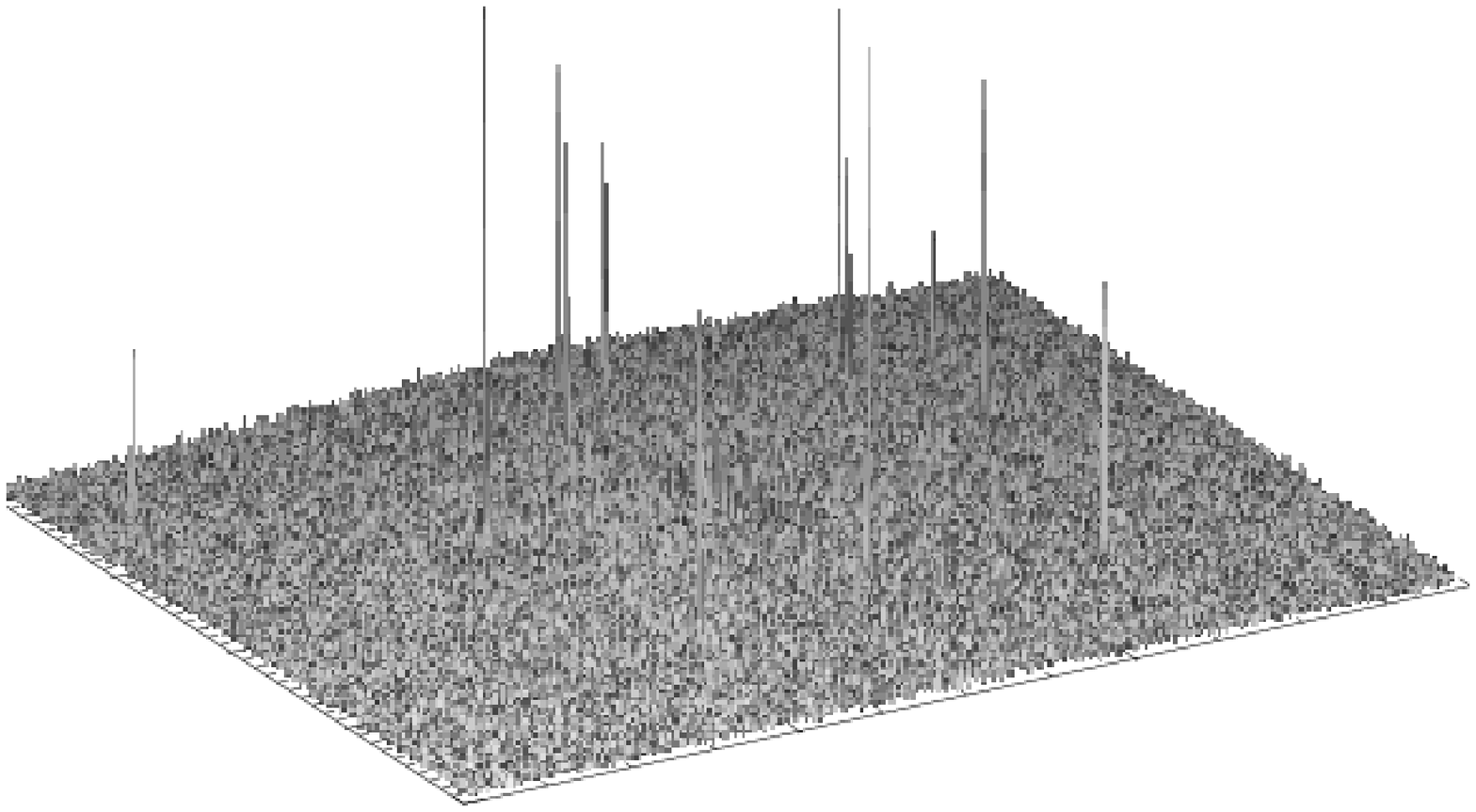}
\caption{(a) Simulated Moire Fringe patterns on the detector plane when
twelve sources are randomly placed. (b) Reconstruction of the three dimensional image of
the photon counts for the twelve sources in a $2^\circ \times 2^\circ$ field.}
\end{figure}

\begin{figure}
\includegraphics[height=5cm]{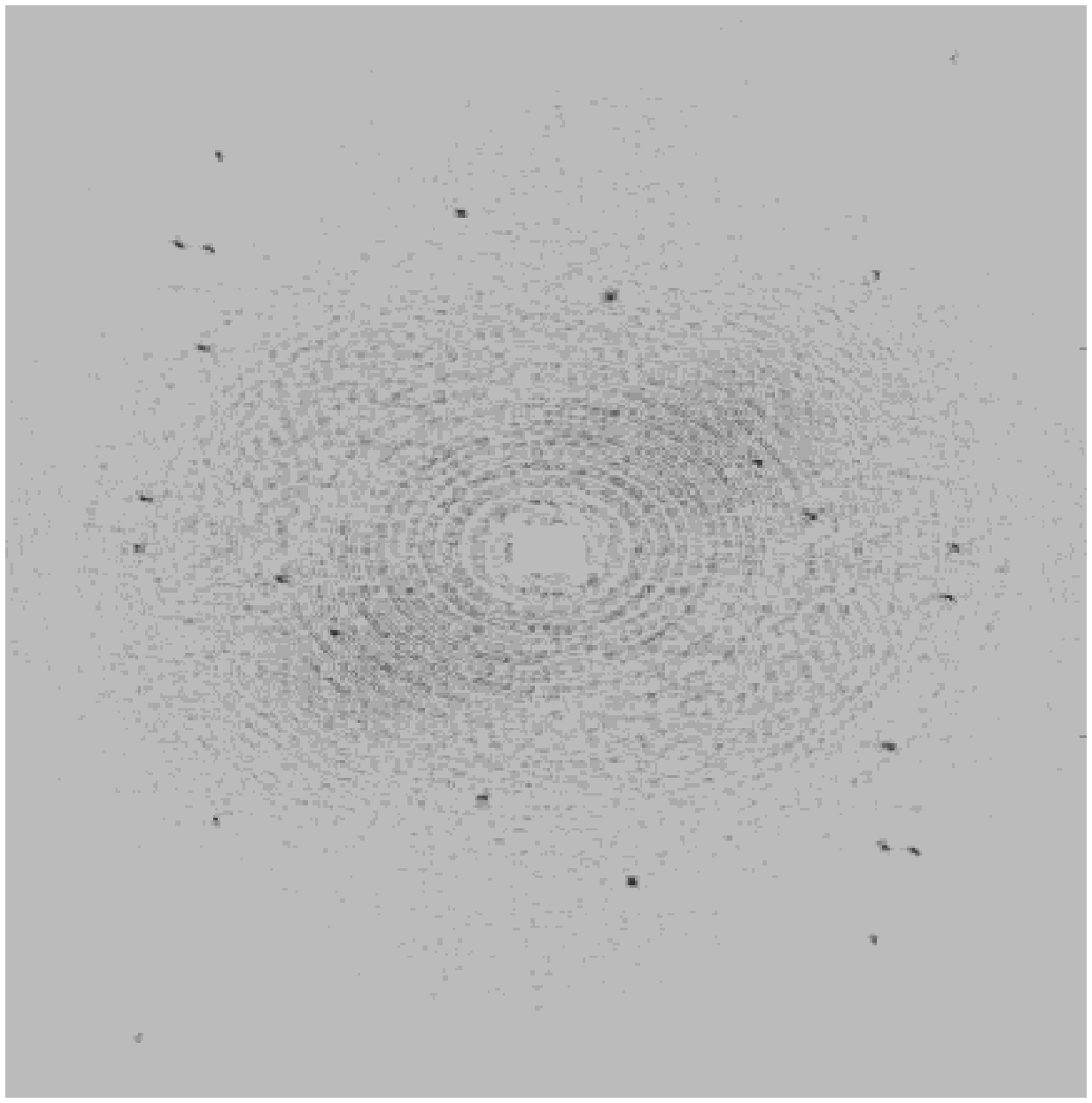}
\hspace{20mm}
\includegraphics[height=5cm]{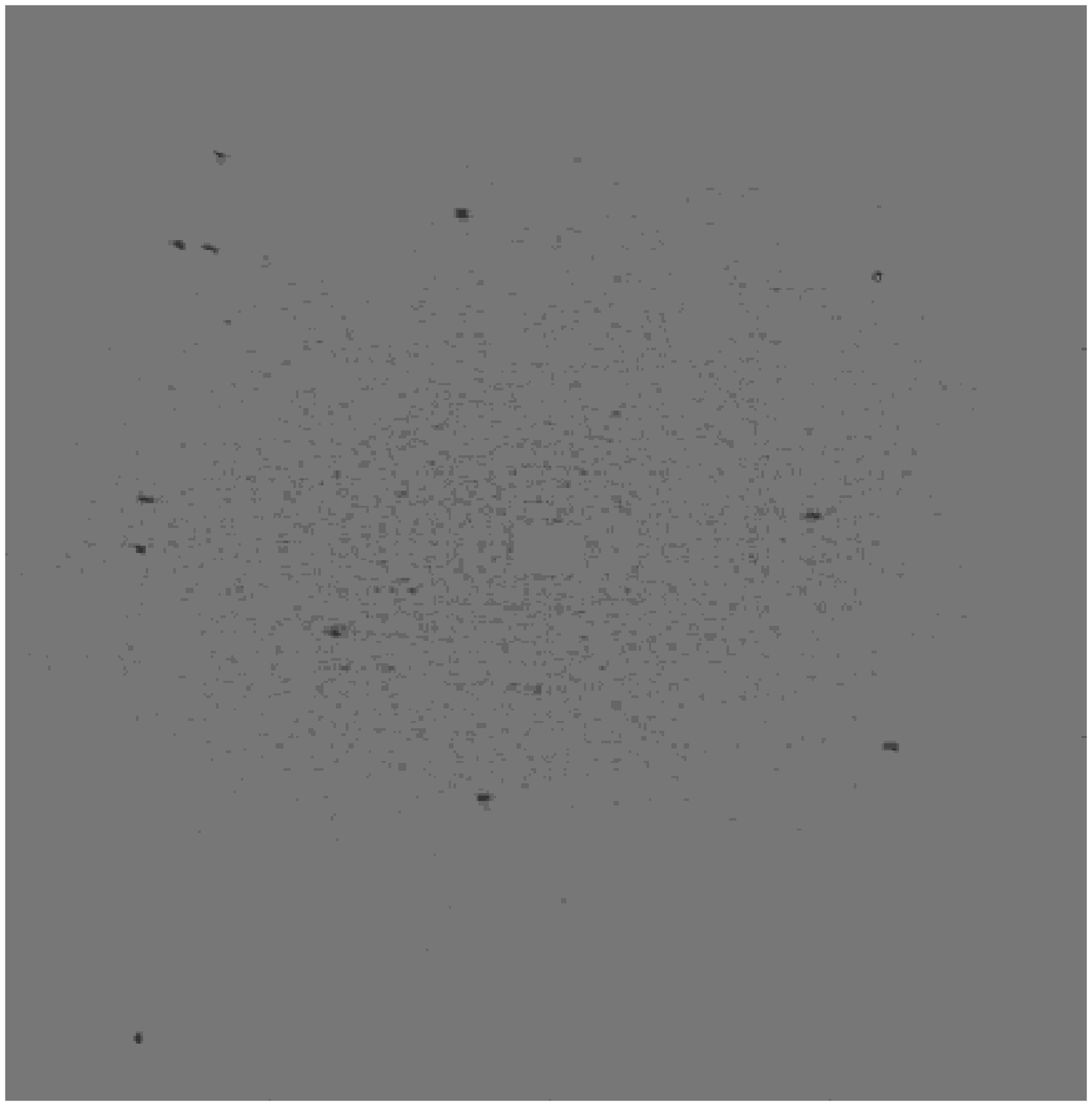}
\caption{(a) Reconstruction of the twelve source locations from the Fig. 4a
using only the Cosine transformation. Note that there are two sets of
twelve sources are symmetrically placed. (b) Same as (a)
except using both the Cosine and and Sine transformations.}
\end{figure}

\section{Concluding remarks}

\noindent
Zone plate combinations are remarkably achromatic high resolution X-ray
imaging devices which are suitable for space applications. We show that very high resolution
may be obtained with suitable choice of distance and zone parameters. 

\begin{theacknowledgments}
We thank  Dr. U. Desai for many helpful discussions. The research of SP
is supported by CSIR/NET scholarship.
\end{theacknowledgments}

\end{document}